\documentclass[aip,jap,showpacs,preprint,floatfix,superscriptaddress]{revtex4-1}
\usepackage[utf8]{inputenc}
\usepackage{graphicx}
\usepackage{dcolumn} 
\usepackage{amsmath,amssymb,bm}
\usepackage{esint}
\usepackage[caption=false]{subfig}
\usepackage{booktabs}
\usepackage{color}

\newcommand{\ternaryx}[3]{#1$_\mathrm{x}$#2$_{1-\mathrm{x}}$#3}
\newcommand{\catplane}[3]{#1$_\mathrm{#3}$#2$_{1-\mathrm{#3}}$}
\newcommand{\ternaryhalf}[3]{#1$_{0.5}$#2$_{0.5}$#3}
\newcommand{\InGaAs}{In$_\mathrm{0.53}$Ga$_{0.47}$As}
\newcommand{\InAlAs}{In$_\mathrm{0.52}$Al$_{0.48}$As}
\newcommand{\wmk}{W/m$\cdot$K}
\begin{document}
\title{Thermal conductivity of ternary III-V semiconductor alloys: the role of mass difference and long-range order}

\author{S. Mei}
\email{song.mei@wisc.edu}
\author{I. Knezevic}
\email{irena.knezevic@wisc.edu}
\affiliation{Department of Electrical and Computer Engineering, University of Wisconsin-Madison, Madison, WI 53706, USA}
\date{\today}

\begin{abstract}
Thermal transport in bulk ternary III-V arsenide (III-As) semiconductor alloys was investigated using equilibrium molecular dynamics with optimized Albe-Tersoff empirical interatomic potentials. Existing potentials for binary AlAs, GaAs, and InAs were optimized to obtain accurate phonon dispersions and temperature-dependent thermal conductivity. Calculations of thermal transport in ternary III-Vs commonly employ the virtual-crystal approximation (VCA),  where the structure is assumed to be a random alloy and all group-III atoms (cations) are treated as if they have an effective weighted-average mass. Here, we showed that is critical to treat atomic masses explicitly, and that the thermal conductivity obtained with explicit atomic masses differs considerably from the value obtained with the average VCA cation mass. The larger the difference between the cation masses, the poorer the VCA prediction for thermal conductivity. The random-alloy assumption in the VCA is also challenged, because X-ray diffraction and transmission electron microscopy show order in InGaAs, InAlAs, and GaAlAs epi-layers. We calculated thermal conductivity for three common types of order [CuPt-B, CuAu-I, and triple-period-A (TPA)] and showed that the experimental results for \InGaAs~ and \InAlAs, which are lattice matched to the InP substrate, can be reproduced in molecular dynamics simulation with 2\% and 8\% of random disorder, respectively. Based on our results, thermal transport in ternary III-As alloys appears to be governed by the competition between mass-difference scattering, which is much more pronounced than the VCA suggests, and the long-range order that these alloys support.
\end{abstract}


\maketitle 
\section{Introduction} \label{sec:intro}
III-V arsenide (III-As) ternary alloys and their superlattices (SLs) are widely used in optoelectronic and thermoelectric devices.\cite{Faist_Science_1994,Kohler_Nature_2002} In particular, the use of III-As alloys brings about great flexibility in the design of the active region of quantum cascade lasers (QCLs).\cite{Razeghi_OME_2013} Despite the wide popularity, even the bulk thermal properties of these ternary alloys have not been completely characterized to date,\cite{Adachi_JAP_2007} which hinders the analysis of device thermal performance.

Measuring the thermal conductivity (TC) of bulk ternary alloys requires growing them strain-free on a substrate. Since GaAs substrate is lattice matched to \ternaryx{Al}{Ga}{As} with any Al composition, TC measurement of AlGaAs were carried out early on, though only at room temperature (RT).\cite{Afromowitz_JAP_1973} Measurements on \ternaryx{In}{Ga}{As} and \ternaryx{In}{Al}{As} are more challenging, as they each only have a specific composition (\InGaAs~and \InAlAs) that is lattice matched to InP, a common substrate for the growth of midinfrared QCLs.\cite{Kim_JEM_2002} As a result, \ternaryx{In}{Ga}{As} and \ternaryx{In}{Al}{As} often exist in pairs inside a strain-balanced SL on top of InP. As the influence that the interfaces in SLs have on thermal transport is poorly understood,\cite{Cahill_JAP_2003,Cahill_APR_2014} it is very difficult to extract the TC of each material from the measurements on SLs. Abrahams \textit{et al.} measured the RT TC of \ternaryx{In}{Ga}{As} with various In compositions, where the samples were formed through zone leveling.\cite{Abrahams_JPCS_1959} However, this method is not applicable to \ternaryx{In}{Al}{As}, as they oxidize easily. Koh \textit{et al.} measured the RT TC of \InGaAs~and \InAlAs~to be 4.02 \wmk~and 2.68 \wmk~, respectively, using both the $3\omega$ technique and time-domain thermoreflectance (TDTR).\cite{Koh_JAP_2009} Sood \textit{et al.} used TDTR to measure the cross-plane TC of \InGaAs/\InAlAs~SLs with different layer thicknesses and extracted the TC of \InGaAs~and \InAlAs~to be 5.0 \wmk~and 1.2 \wmk, respectively. The two sets of results disagree, with the disparity being particularly pronounced for \InAlAs. Furthermore, both sets of results were obtained for \InGaAs~and \InAlAs~lattice matched to InP. We still have very limited knowledge about the TC of \ternaryx{In}{Ga}{As} and \ternaryx{In}{Al}{As} bulk materials with various compositions.

Our recent theoretical work solved the Boltzmann transport equation for phonons under the relaxation-time approximation in order to describe the TC of III-As semiconductors.\cite{Mei_JAP_2015} We accurately captured the TC of binary compounds (AlAs, GaAs, and InAs) from 100 K to 600 K with full acoustic phonon dispersions. Additionally, we adopted the virtual crystal approximation\cite{Abeles_PR_1963} (VCA) to compute the TC of ternary alloys \ternaryx{Al}{Ga}{As}, \ternaryx{In}{Ga}{As}, and \ternaryx{In}{Al}{As} at RT. For  \ternaryx{In}{Ga}{As} and \ternaryx{Al}{Ga}{As}, our calculated TC agreed  reasonably well with experimental values with various $x$ compositions.\cite{Abrahams_JPCS_1959,Adachi_JAP_1983,Afromowitz_JAP_1973} For \ternaryx{In}{Al}{As}, an experimental TC value was only available for \InAlAs.\cite{Koh_JAP_2009,Sood_APL_2014} Our calculated TC is 2.82 \wmk, which is close to the Koh \textit{et al.} value of 2.68 \wmk,\cite{Koh_JAP_2009} but far from the Sood \textit{et al.} value of 1.2 \wmk.\cite{Sood_APL_2014} Considering that we have benchmarked the technique for binaries, we speculate that the VCA that is not entirely suitable for characterizing the TC of ternary III-As alloys.

Within the VCA, there are two major assumptions whose validity needs to be questioned when it comes to the calculation of TC in ternary alloys \ternaryx{A}{B}{C}. (1) All cations can effectively be replaced with an effective averaged cation, whose mass is calculated as the weighted average of cation masses, \textit{i.e.},  m$_\mathrm{III}=x\mathrm{m}_\mathrm{A}+(1-x)\mathrm{m}_\mathrm{B}$. (2) The alloy is random, \textit{i.e.}, cation sites are randomly taken by atom A or atom B, with the frequency proportional to cation abundance [$x$ and $(1-x)$, respectively]. Although an effective mass-difference scattering rate is often employed in Boltzmann-equation-based approaches to compensate for the fact that assumption (1) eliminates scattering caused by cation mass difference,\cite{Abeles_PR_1963} these rates are calculated under the assumption  that the cation mass difference is small with respect to the average cation mass. Therefore, the validity of this perturbative mass-difference-scattering approach becomes suspect  when the cation masses are very different. In our case, m$_\mathrm{Al}=26.98$ au, m$_\mathrm{Ga}=69.72$ au, and m$_\mathrm{In}=114.82$ au, so it makes sense that the model would work better for \ternaryx{In}{Ga}{As} and \ternaryx{Al}{Ga}{As} than it does for \ternaryx{In}{Al}{As}, since the mass difference between Al atoms and In atoms is significant. The second assumption is directly contradicted by a number of X-ray and transmission-electron-microscopy (TEM) experiments conducted on ternary III-V alloys grown epitaxially on GaAs or InP substrate.\cite{Shahid_PRL_1987,Shin_MT_2006,Kuan_APL_1987,Mori_ASS_2004,Kulik_MRSSP_2000,Forrest_JMR_2000,Suzuki_2002_Spontaneous,Suzuki_APL_1998,Gomyo_PRL_1994,Ohkouchi_ASS_2005,Kuan_PRL_1985,Bernard_APL_1990} These experiments show that group III atoms in ternary III-As epi-layers are arranged with a certain order rather than completely randomly. This ordered structure is also backed up by an observable change in the width of the electronic band gap of these materials. The most commonly observed types of ordering for \InGaAs~and \InAlAs~are the CuPt-B order\cite{Shahid_PRL_1987,Shin_MT_2006,Kuan_APL_1987,Kulik_MRSSP_2000} and the triple-period-A (TPA) order.\cite{Mori_ASS_2004,Forrest_JMR_2000,Suzuki_2002_Spontaneous,Suzuki_APL_1998,Gomyo_PRL_1994,Ohkouchi_ASS_2005} The CuAu-I order is more common in \ternaryhalf{Al}{Ga}{As}\cite{Kuan_PRL_1985,Bernard_APL_1990}, but could also be observed in \InGaAs~or \InAlAs~under appropriate growth conditions.\cite{Suzuki_2002_Spontaneous,Bernard_APL_1990}

In this work, we used molecular dynamics (MD) to study thermal transport in ternary III-As alloys. In a MD simulation, the mass and location of each atom are tracked in real space; therefore, it is straightforward to include individual atom masses and the exact alloy structure explicitly in the simulation. We adopted the Tersoff\cite{Tersoff_PRB_1988,Tersoff_PRB_1989} empirical interatomic potentials (EIPs) to describe the interaction between atoms. The parameters we used in the EIPs were obtained by an optimization technique, starting from the existing values,\cite{Sayed_NIMPRB_1995,Powell_PRB_2007,Hammerschmidt_PRB_2008} with the goal of better capturing the phonon properties of binary III-As semiconductors.\cite{Lindsay_PRB_2010a} We then used the optimized EIPs to study the influence of mass difference and ordering on the TC of ternary III-As alloys. The results show that the atom mass must be explicitly considered, especially in In-based compounds. Ordering can significantly increase the TC and partial ordering is a likely reason for the higher experimentally obtained thermal conductivities than what the VCA within the Boltzmann transport framework predicts. Moreover, TC can vary a great deal with the introduction of even low levels of disorder.

In Sec.~\ref{sec:optimization}, we discuss the way we compute the bulk TC from an MD simulation and our process of optimizing the Tersoff potentials for AlAs, GaAs, and InAs. In Sec.~\ref{sec:mass}, we apply the optimized potentials to ternary alloys. We isolate the effect of mass difference by comparing the TC obtained from structures where each cation has its own mass and those where all cations have the effective VCA mass. In Sec.~\ref{sec:order}, we study the TC of bulk ternary III-As alloys with perfect ordering and lattice-matched \InGaAs~ and \InAlAs~that contain small levels of random disorder. Finally, we summarize our findings in Sec.~\ref{sec:conclusion}.

\section{Thermal Conductivity from Equilibrium Molecular Dynamics} \label{sec:EMD_TC}
\subsection{The Green-Kubo Formula for Thermal Conductivity}\label{subsec:GK}
All the MD simulations in this work were carried out in the LAMMPS\cite{Plimpton_JCP_1995} package. We used equilibrium molecular dynamics (EMD) together with the Green-Kubo (GK) formula to compute the TC of compound semiconductors. When a system is equilibrated and then kept at a set temperature, the TC can be derived from the fluctuation--dissipation theorem using instantaneous heat flux.\cite{Frenkel_2001_Understanding} In a three-dimensional (3D) bulk material where all three directions are equivalent, thermal conductivity $\kappa$ is described by the GK formula as\cite{Frenkel_2001_Understanding}
\begin{equation}\label{equ:GK}
\kappa=\frac{1}{3k_\mathrm{B}VT^2}\int_{0}^{\infty}\langle\mathbf{S}(t)\mathbf{S}(0)\rangle dt,
\end{equation}
where $k_\mathrm{B}$ is the Boltzmann constant, $V$ is the system volume, and $T$ is the system temperature. $\mathbf{S}(t)$ is the instantaneous heat flux calculated from the atom velocities and local potentials. The integral is over the heat-current autocorrelation function,  $\langle\mathbf{S}(t)\mathbf{S}(0)\rangle$. In an MD simulation, the time is discretized and the integration in Equation~(\ref{equ:GK}) is converted to a sum.

For cubic bulk semiconductors, the TC is expected to be isotropic. As a result, we used cubic simulation cells with periodic boundary conditions applied in all directions. In a typical simulation, the system was first initialized at the desired temperature $T$ by assigning each atom a random velocity that follows the thermal distribution at $T$. Then the system was equilibrated as an $NPT$ ensemble (constant number of atoms, constant pressure at 1 atm, and constant temperature $T$) using the Nos$\acute{\mathrm{e}}$-Hoover barostats and thermostats\cite{Nose_JCPS_1984,Hoover_PRA_1985} for 100 ps. After that, the system was further equilibrated as an $NVE$ ensemble (constant number of atoms, constant volume, and constant system energy) for another 100 ps before the heat flux is collected. The instantaneous heat flux was calculated and output into a file at every time step, for 5 million steps. A script was written to post-process the output file and obtain the heat-current autocorrelation function and its running integral. We made sure the integral saturated and extracted the bulk TC according to Equation~(\ref{equ:GK}). The time step was chosen to be small enough to ensure the system stayed stable throughout the simulation -- here, the time step was 1 fs for binary materials and 0.1 fs for ternary alloys. For each simulation (given material and temperature), several random starting velocity distributions were used and the final result was averaged among the different runs. We tested the simulation domain sizes to make sure there was no size effect. The final results for binary III-As were obtained with a simulation cell $10a_0\times10a_0\times10a_0$ in size, where $a_0$ is the lattice constant for the material. For ternary alloys, $8a_0\times8a_0\times8a_0$ is enough for the TC to converge.  We used $9a_0\times9a_0\times9a_0$ cells for alloys with TPA ordering, as the algorithm for generating cells requires that the system size be a multiple of 3 (more detail in Sec.~\ref{sec:order}).
\subsection{Quantum Correction of Temperature}\label{subsec:qcT}
\begin{figure}
	\centering
	\includegraphics[width=0.7\columnwidth]{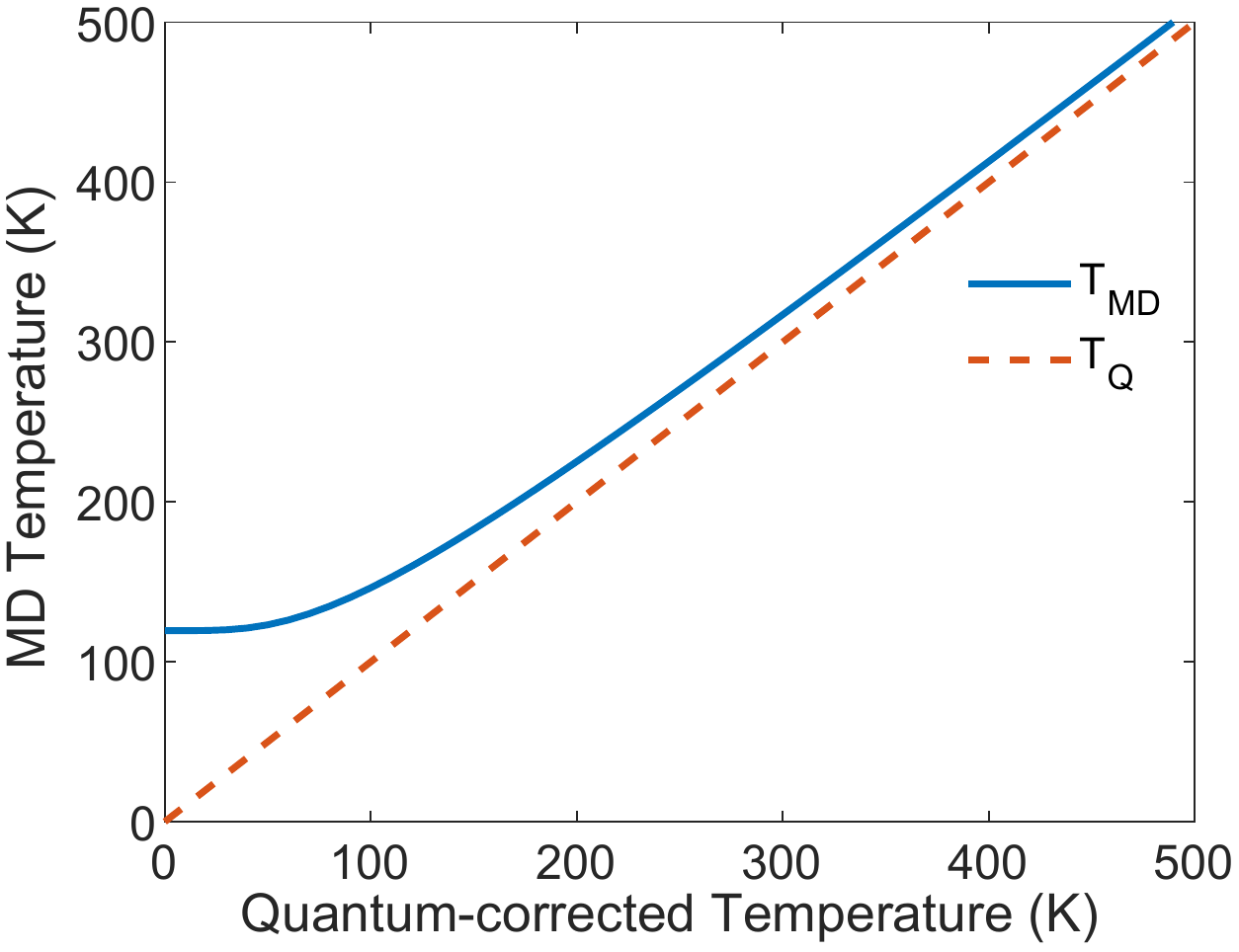}
	\caption{A typical mapping between the MD-calculated temperature $T_{\mathrm{MD}}$ and the quantum-corrected temperature $T_{\mathrm{Q}}$ (figure shown for GaAs). $T_{\mathrm{MD}}$ and $T_{\mathrm{Q}}$ coincide at high temperatures, but differ a great deal at low temperatures.}
	\label{fig:qc}
\end{figure}

In MD simulations, system temperatures are calculated following the rules of classical statistical mechanics.~\cite{Wang_PRB_1990} We adopted a simple quantum-correction procedure for the temperatures by mapping the kinetic energy of an MD system at temperature $T_{\mathrm{MD}}$ onto that of a quantum system with temperature $T_{\mathrm{Q}}$\cite{Wang_PRB_1990,Lukes_JHT_2007}
\begin{equation}\label{equ:Tcorrection}
\frac{3}{2}Nk_\mathrm{B}T_{\mathrm{MD}}=\sum_{\mathrm{b}}\sum_\mathbf{k}\hbar\omega(\mathbf{k},\mathrm{b})\left\{\frac{1}{2}+\frac{1}{\exp\left[\frac{\hbar\omega(\mathbf{k},\mathrm{b})}{k_\mathrm{B}T_{\mathrm{Q}}}\right]+1}\right\}.
\end{equation}
On the left-hand side, $N$ is the number of atoms in the system and $k_\mathrm{B}$ is the Boltzmann constant. On the right-hand side, the summation is over all the phonon branches $\mathrm{b}$ and wave vectors $\mathbf{k}$. $\hbar\omega(\mathbf{k},\mathrm{b})$ is the corresponding phonon energy. For the right-hand side, we used an approximate isotropic phonon dispersion fitted to the full phonon dispersion, based on our previous work.\cite{Mei_JAP_2015} Figure~\ref{fig:qc} shows the mapping between the quantum-corrected temperature and the MD temperature for GaAs (the curve is similar for AlAs and InAs) between 0 K and 500 K. We see that $T_{\mathrm{MD}}$ and  $T_{\mathrm{Q}}$ coincide at higher temperatures, but differ a great deal at lower temperatures. At room temperature, there is typically a $4\%-8\%$ difference between $T_{\mathrm{MD}}$ and $T_{\mathrm{Q}}$ for III-As. Henceforth, all the temperatures listed in this work are the quantum-corrected temperatures $T_{\mathrm{Q}}$.

\section{Optimizing the Tersoff Potentials} \label{sec:optimization}
\subsection{Tersoff Potentials}\label{subsec:tersoff}
The Tersoff EIP was proposed by Tersoff in 1988 for silicon,\cite{Tersoff_PRB_1988} extended to SiC in 1989,\cite{Tersoff_PRB_1989} and later successfully parameterized for most III-V binary compounds.~\cite{Sayed_NIMPRB_1995,Nordlund_CMS_2000,Powell_PRB_2007,Hammerschmidt_PRB_2008}
Tersoff potentials have a cutoff distance that limits the atom interactions to nearest neighbors, which is an advantage in our case: the nearest-neighbor interaction makes it easy to apply the optimized potentials for binary III-As to ternary III-As alloys.

In the LAMMPS package, the Tersoff potential is described using 14 parameters in the following form.
The total energy of the system is the summation of energy between each pair:
\begin{equation}
	E=\frac{1}{2}\sum_i\sum_{j\neq i}V_{ij}.
\end{equation}
\begin{subequations}
The pair interaction potential $V_{ij}$ between atom $i$ and atom $j$ is described by the competition between a repulsive term $f_R$ and an attractive term $f_A$, and is modulated by a cutoff term $f_C$ so that only nearest-neighbor interactions are included:
\begin{equation}
	V_{ij}=f_C(r_{ij})\left[f_R(r_{ij})+b_{ij}f_A(r_{ij})\right].
\end{equation}
The cutoff function is
\begin{equation}\label{equ:cutoff}
f_C(r)=
\begin{cases}
1 &:r<R-D,\\
\frac{1}{2}-\frac{1}{2}\sin\left(\frac{\pi}{2}\frac{r-R}{D}\right)&:R-D<r<R+D,\\
0 &:r>R+D,
\end{cases}
\end{equation}
where $r$ is the variable. From Equation~(\ref{equ:cutoff}), the cutoff length is $R+D$ and there is a transition window with width $2D$. Both $R$ and $D$ are parameters of the Tersoff potential.
Both the repulsive and the attractive terms have an exponential form with $A,~B,~\lambda_1$ and $\lambda_2$ being the potential parameters:
\begin{equation}\label{equ:rep}
	f_R(r)=A\exp(-\lambda_1r),
\end{equation}
\begin{equation}\label{equ:att}
	f_A(r)=-B\exp(-\lambda_2r).
\end{equation}
To include the influence of the bond angle and length, the attraction term $f_A$ is further modified by the bond angle term $b_{ij}$ where
\begin{equation}
	b_{ij}=\left(1+\gamma^n\xi_{ij}^n\right)^{-\frac{1}{2n}},
\end{equation}
and
\begin{align}
\xi_{ij}&=\sum_{k\neq i,j}f_C(r_{ik})g(\theta_{ijk})\exp\left[\lambda_3^m\left(r_{ij}-r_{ik}\right)^m\right],\\
g(\theta)&=\delta_{ijk}\left(1+\frac{c^2}{d^2}-\frac{c^2}{d^2+(\cos\theta-\cos\theta_0)^2}\right).
\end{align}
The summation in $\xi_{ij}$ goes over all the central atom $i$'s neighbors within the cutoff except for the neighbor $j$ whose interaction with $i$ we are considering. The bond angle $\theta_{ijk}$ can be calculated with the three atoms' coordinates and $\delta_{ijk}$ is a scaling parameter that is unity in most cases. $\gamma$, $n$, $\lambda_3$, and $m$ are potential parameters. In particular, $m$ can only take the value of 1 or 3. $c,~d,$ and $\cos\theta_0$ are bond-angle-related potential parameters. Note that $\cos\theta_0$ is often denoted as $h$; the form of a cosine simply reminds us this parameter can only take values between -1 and 1.

A variation of the original Tersoff potential is the Albe-Tersoff potential where the strength and the decay rate in Equation~(\ref{equ:rep}) and (\ref{equ:att}) are expressed through parameters $D_e,~S,~\beta,$ and $r_e$ in place of $A,~B,~\lambda_1$, and $\lambda_2$.\cite{Albe_PRB_2002} The relationship between the two sets of parameters can be derived as
\begin{align}
	A &=\frac{D_e}{S-1}\exp(\beta\sqrt{2Sr_e}),\\
	B &=\frac{SD_e}{S-1}\exp(\beta\sqrt{2S/r_e}),\\
	\lambda_1 &=\beta\sqrt{2Sr_e},\\
	\lambda_2 &=\beta\sqrt{2S/r_e},
\end{align}
In the case of Albe-Tersoff potential, we work with the parameters $D_e,~S,~\beta,$ and $r_e$ and then use a script to convert them to $A,~B,~\lambda_1$, and $\lambda_2$ for the input potential file of LAMMPS.
\end{subequations}

\subsection{Phonon Dispersion from EIP}\label{subsec:dispersion}
With a given EIP, the phonon frequencies $\omega(\mathbf{k},\mathrm{b})$ for wave vector $\mathbf{k}$ and branch $\mathrm{b}$ can be obtained by diagonalizing the dynamical matrix\cite{Srivastava_1990_Physics}
\begin{equation}\label{equ:DM}
	\mathcal{D}_{\alpha\beta}(ij|\mathbf{k})=\frac{1}{\sqrt{m_im_j}}\sum_{l'}\phi_{\alpha\beta}(0i;l'j)e^{i\mathbf{k}\mathbf{x}(l')},
\end{equation}
where $l'j$ represents atom $j$ in the $l'$th unit cell and the summation is over all the relevant neighbors of the central atom $i$. $\mathbf{x}(l')$ is the relative location of unit cell $l'$ with respect to the unit cell that atom $i$ is in. The interatomic force constants (IFCs) between atoms $i$ and $j$, denoted by $\phi_{\alpha\beta}(0i;l'j)$, were calculated using centred finite differences:
\begin{equation}
	\phi_{\alpha\beta}^{ij}=\frac{\partial V_{ij}^2}{\partial\alpha\partial\beta}=\frac{V(\mathrm{r}_{ij}+d_\alpha+d_\beta)-V(\mathrm{r}_{ij}-d_\alpha+d_\beta)-V(\mathrm{r}_{ij}+d_\alpha-d_\beta)+V(\mathrm{r}_{ij}-d_\alpha-d_\beta)}{4d_\alpha d_\beta}.
\end{equation}

\noindent Phonon dispersion curves were obtained by computing the phonon frequencies for multiple $\mathbf{k}$s along a certain direction in the first Brillouin zone (1BZ). The sound velocities were obtained from the acoustic branches near the zone center ($\Gamma$ point) as
\begin{equation}
	|\mathbf{v}_{s,\mathrm{b}}|=|\nabla_\mathbf{k}\omega(\mathbf{k},\mathrm{b})|,
\end{equation}
where the subscript $\mathrm{b}$ is the branch index. Since the phonon dispersion is isotropic near the $\Gamma$ point, we used the dispersion curves along the [100] ($\Gamma-\mathrm{X}$) direction to calculate the sound velocities. The two transverse acoustic (TA) branches are degenerate along the $\Gamma-\mathrm{X}$ direction, so we only need one scalar velocity for the TA branches ($v_{s,\mathrm{TA}}$) and one for the longitudinal acoustic (LA) branch ($v_{s,\mathrm{LA}}$).

\subsection{Parameter Optimization}\label{subsec:optimization}
Most Tersoff potentials are parameterized so as to accurately capture the mechanical properties of materials.\cite{Tersoff_PRB_1988,Tersoff_PRB_1989,Sayed_NIMPRB_1995,Nordlund_CMS_2000,Powell_PRB_2007,Hammerschmidt_PRB_2008} However, for the EIPs to be good at describing thermal transport, they also must produce good phonon properties. Lindsay and Broido\cite{Lindsay_PRB_2010a} optimized the Tersoff potential of carbon (C) for thermal transport. They used a $\chi^2$-minimization, where
\begin{equation}\label{equ:chisq}
\chi^2=\sum_i\frac{\left(\eta_i-\eta_{\mathrm{exp},i}\right)^2}{\eta_{\mathrm{exp},i}^2}\zeta_i,
\end{equation}
and where $i$ runs over all the physical properties they optimize for. $\eta_i$ and $\eta_{\mathrm{exp},i}$ are, respectively, the calculated and experimental values of the physical property $i$. $\zeta_i$ is the weighting factor determining the relative importance of each physical property in the optimization process. They assigned the most weight to the phonon frequencies and sound velocities, because of their crucial rule in thermal transport.\cite{Lindsay_PRB_2010a}

Here, we followed a similar minimization approach to improve Tersoff EIPs  for  describing thermal transport in III-Vs. The physical properties we optimized for are the lattice constant $a_0$, the cohesive energy $E_\mathrm{coh}$, and sound velocities $v_{s,\mathrm{TA}}$ and $v_{s,\mathrm{LA}}$. Note that the accuracy  of phonon dispersions only guarantees the accuracy of the second derivatives of the EIPs (or the IFCs) (Sec.~\ref{subsec:dispersion}), while the finite TC of crystals originates from the phonon--phonon scattering, related to the third and higher derivatives of the EIPs. Since the strength of these higher-order interactions is not easily accessible in experiments, we used the temperature-dependent TC measured in experiments as an additional gauge, because the temperature variation of TC is dictated by phonon-phonon interactions. First, we optimized the EIPs to match the phonon dispersion following Lindsay and Broido.\cite{Lindsay_PRB_2010a} Then we used the optimized potential to calculate the temperature-dependent TC, following the procedure described in Sec.~\ref{sec:EMD_TC}. We were most interested in the temperature range between 100 K and 500 K, because most devices operate in this range. A comparison between the calculated and measured TC instructed us to further adjust the weights in the optimization process. This process was repeated until we obtained a satisfactory temperature-dependent TC from the potentials.

For simplicity, we used existing parameterizations as starting points. To choose the best starting point, we calculate the TDTC with existing potentials. For the three binary III-As materials (AlAs, GaAs, and InAs) we are most interested in, the potentials that yield the temperature-dependent TC closest to experiment are from Sayed \textit{et al.}\cite{Sayed_NIMPRB_1995}, Powell \textit{et al.}\cite{Powell_PRB_2007}, and Hammerschmidt \textit{et al.}\cite{Hammerschmidt_PRB_2008}, respectively. Incidentally, all three sets are parameterized in the Albe-Tersoff form. Like Lindsay and Broido,\cite{Lindsay_PRB_2010a} we tried to adjust just a few parameters. $R$ and $D$ were left untouched, because they dictate the cutoff of the EIP. $m$ also stayed constant, as per LAMMPS's requirement. Among the remaining parameters, we found that $D_e,~\beta,~c,~d,$ and $h$ are very effective in adjusting the four physical properties we want to optimize. Therefore, during the $\chi^2$ minimization, we only varied these five parameters.

\subsection{Optimized Potentials}\label{subsec:optpot}
In the optimization, we found that the parameters for GaAs from Powell \textit{et al.}\cite{Powell_PRB_2007} yielded very good sound velocities as well as temperature-dependent TC from 100 K to 500 K. Therefore, we adopted this set of parameters as it was. However, the parameters for AlAs\cite{Sayed_NIMPRB_1995} and InAs\cite{Hammerschmidt_PRB_2008} both had to be optimized. Table~\ref{tab:optparameters} shows the optimized parameters for AlAs and InAs. As mentioned in Sec.~\ref{subsec:optimization}, the parameters other than the five listed were kept the same as in the original sets.
\begin{table}
	\begin{tabular}{c@{\hspace{2em}}|@{\hspace{2em}}c}
		\hline\hline
		AlAs & InAs \\
		\hline
		$D_e=2.6372$ & $D_e=1.9949$\\
		$\beta=1.6948$ & $\beta=1.7660$\\
		$c = 1.4145$ & $c = 4.0249$\\
		$d=0.9116$ & $d=1.0157$\\
		$h=-0.6172$ & $h=-0.6096$\\
		\hline
	\end{tabular}
	\caption{Optimized parameters for the Albe-Tersoff potentials for AlAs and InAs. Parameters not listed are the same as in the original sets.\cite{Sayed_NIMPRB_1995,Hammerschmidt_PRB_2008}}
	\label{tab:optparameters}
\end{table}

\begin{figure}
	\centering
	\includegraphics[width=3.3 in]{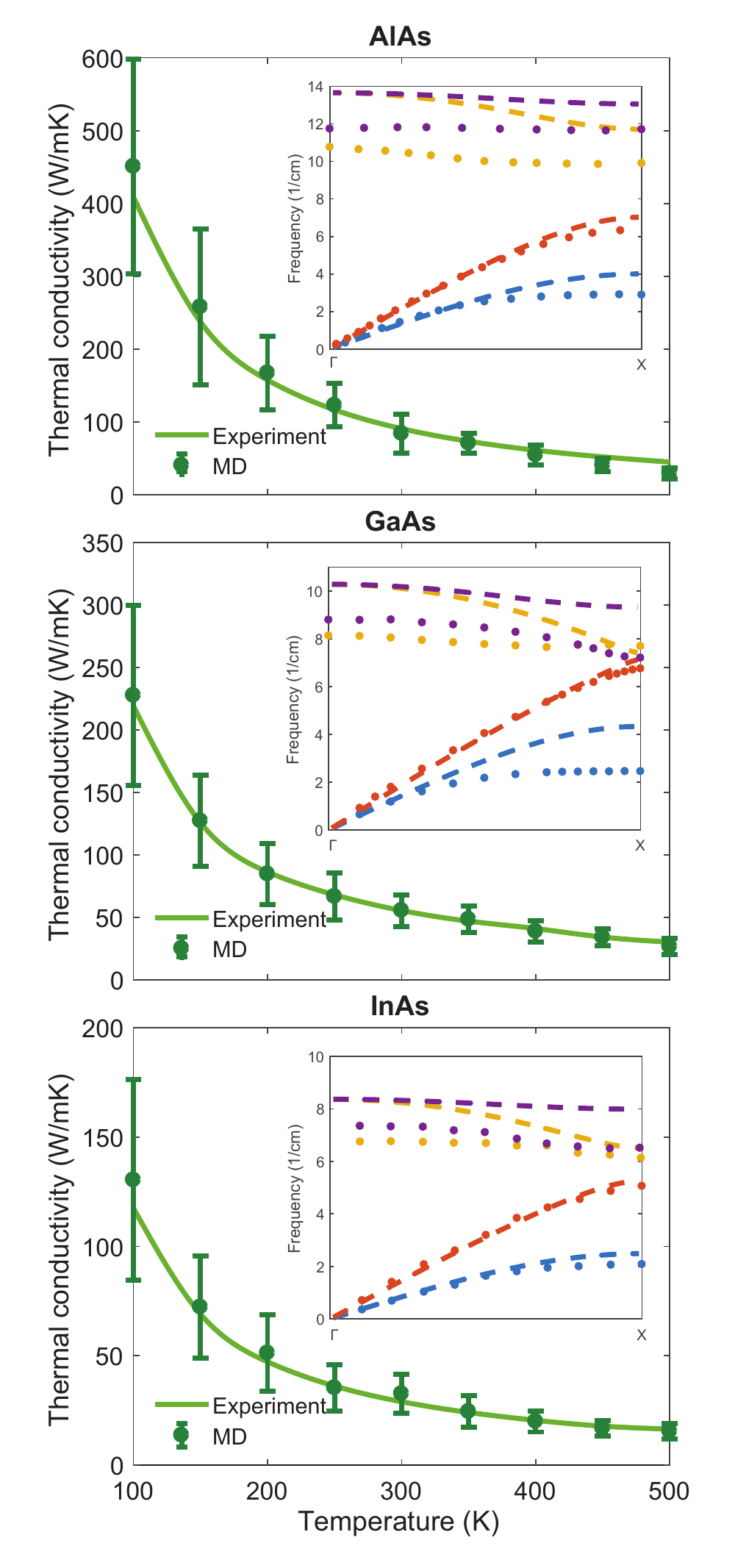}
	\caption{Temperature-dependent thermal conductivity, as calculated in this work with optimized potentials (dark green dots with error bars) and from experiment (light green solid lines) for AlAs,\cite{Evans_JQE_2008} GaAs,\cite{Inyushkin_SST_2003,Amith_PR_1965} and InAs.\cite{Tamarin_SPS_1971,Guillou_PRB_1972,Bowers_JAP_1959} The insets show the calculated phonon dispersions (dashed lines) along the $\Gamma-\mathrm{X}$ direction along with experimental values (dots).\cite{Lindsay_PRB_2013,Dorner_JPCM_1990,Orlova_PSSB_1983,Carles_PRB_1980} }
	\label{fig:tdtc}
\end{figure}

The three panels in Fig.~\ref{fig:tdtc} show the calculated temperature-dependent TC (dark green dots with error bars) in comparison with experimental data~\cite{Inyushkin_SST_2003,Amith_PR_1965,Evans_JQE_2008,Tamarin_SPS_1971,Guillou_PRB_1972,Bowers_JAP_1959} (light green solid lines) for AlAs, GaAs, and InAs, respectively. The insets show the calculated phonon dispersions along the $\Gamma-\mathrm{X}$ direction (dashed lines) in comparison with measured values~\cite{Lindsay_PRB_2013,Dorner_JPCM_1990,Orlova_PSSB_1983,Carles_PRB_1980} (dots).

\section{Thermal Conductivity of Ternary Alloys: The Role of Cation Mass Difference} \label{sec:mass}
After obtaining the optimized EIPs for binary compounds, we applied these potentials to ternary alloys. We investigated the validity of the first assumption in the VCA (all cation atoms are assigned the same weighted-average mass) in predicting the TC. For this part, we kept the second assumption that all ternary alloys were random alloys. We considered \ternaryx{In}{Ga}{As} and \ternaryx{In}{Al}{As} alloys with $x$ varying from 0.1 to 0.9. For each $x$, we generated 10 different random simulation cells with the corresponding composition. For each random configuration, we conducted 20 simulations with different starting velocity distributions. Each of the final TC value was  averaged over the 200 runs.

\begin{figure}
	\centering
	\includegraphics[width=0.7\columnwidth]{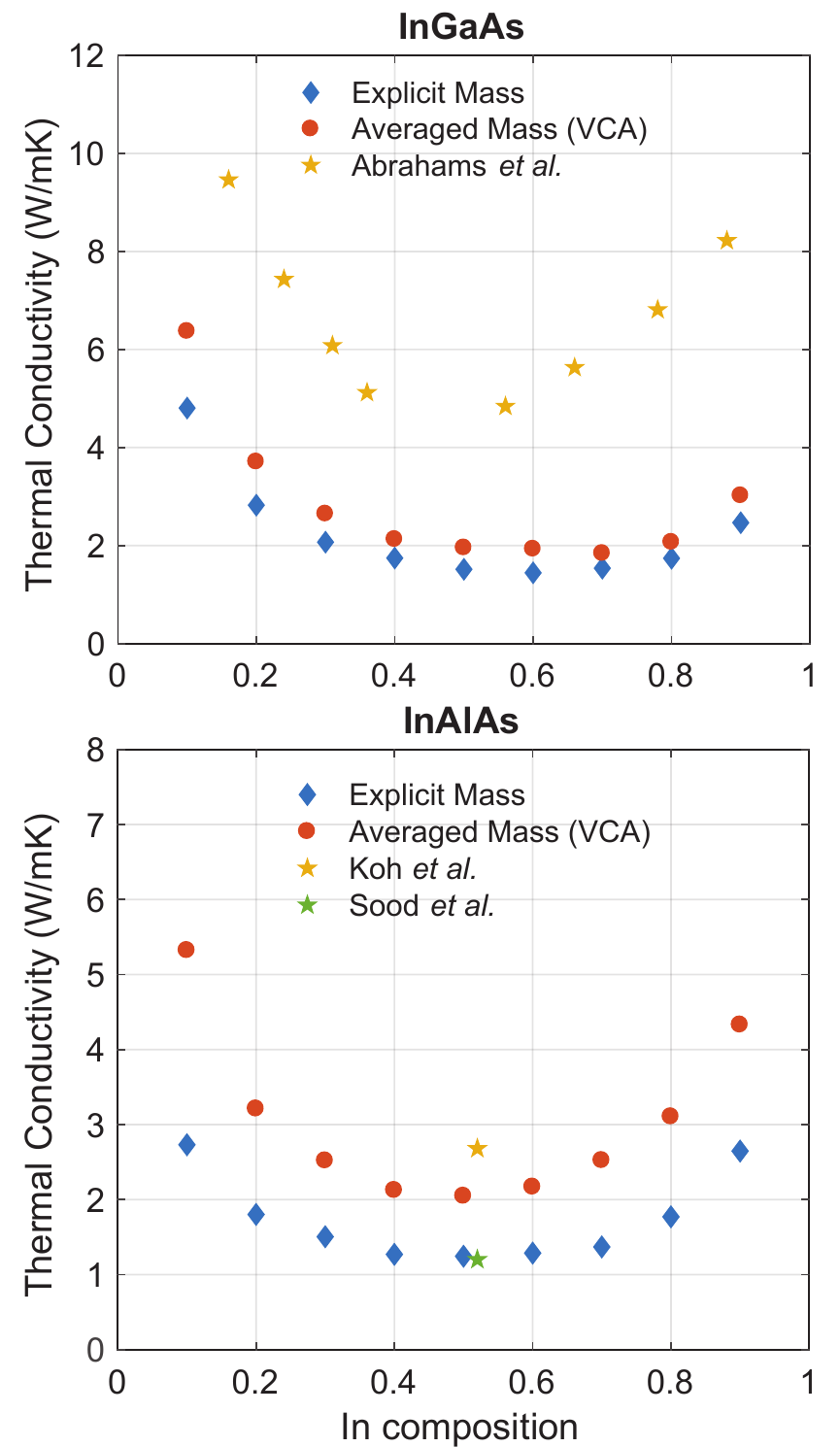}
	\caption{Thermal conductivity as a function of In composition in \ternaryx{In}{Ga}{As} and \ternaryx{In}{Al}{As} random alloys at room temperature. In blue diamonds, we show the calculated TC for the case where each atom's mass is included explicitly. Results from the case where all cations take on the averaged (VCA) mass are shown in red dots. Stars show experimental data. \cite{Abrahams_JPCS_1959,Koh_JAP_2009,Sood_APL_2014}}
	\label{fig:randAlloy_TC}
\end{figure}

In order to isolate the effect of mass difference, we carried out two sets of simulations (all at RT). In the first set, all the cation atoms kept their own masses (m$_\mathrm{Al}=26.98$ au, m$_\mathrm{Ga}=69.72$ au, and m$_\mathrm{In}=114.82$ au). In the second set, the cations were all assigned the same weighted-average VCA mass $\mathrm{m}_{\mathrm{avg}}=x\mathrm{m}_{\mathrm{In}}+(1-x)\mathrm{m}_{\mathrm{Ga/Al}}$. Other than this difference, the two sets used the same EIPs and random (not ordered) configurations in the simulation. Figure~\ref{fig:randAlloy_TC} shows the TC data for both \ternaryx{In}{Ga}{As} and \ternaryx{In}{Al}{As}. The results from the explicit-mass (EM) case and the averaged-mass VCA case are shown in blue diamonds and red dots, respectively. Existing experimental data are shown in stars, for comparison.\cite{Abrahams_JPCS_1959,Koh_JAP_2009,Sood_APL_2014}

From Fig~\ref{fig:randAlloy_TC}, the EM TC is consistently lower than the VCA  TC across the range of different In\% for both \ternaryx{In}{Ga}{As} and \ternaryx{In}{Al}{As}. Moreover, the difference between the EM and VCA is much more pronounced in \ternaryx{In}{Al}{As} (where the cation mass difference is larger) than in \ternaryx{In}{Ga}{As} (where the cation mass difference is smaller). Therefore, it is evident that mass-difference scattering of lattice waves is critical in III-V ternary random alloys. Using an averaged VCA mass for cations will underestimate the mass-difference scattering; the larger the mass difference, the greater the underestimate. Additionally, in the case of \ternaryx{In}{Ga}{As}, both the EM TC and the VCA TC are significantly lower  than the experimental value; in the case of \ternaryx{In}{Al}{As}, the two experimental results are so far apart that the EM TC and AM TC fall in between the two measurements. As a result, we can conclude that there must be another mechanism that competes with mass-difference scattering to influence the TC of ternary alloys.

\section{Thermal Conductivity of Ternary Alloys: The Role of Order} \label{sec:order}
Experimentalists who grow and characterize III-V epitaxial layers discovered long ago that both long-range and short-range order exists in ternary III-V alloys.\cite{Shahid_PRL_1987,Kuan_APL_1987,Mori_ASS_2004,Kulik_MRSSP_2000,Forrest_JMR_2000,Suzuki_2002_Spontaneous,Suzuki_APL_1998,Suzuki_MRS_1999,Gomyo_PRL_1994,Ohkouchi_ASS_2005,Kuan_PRL_1985,Bernard_APL_1990} Order leads to changes in bond length and electronic band gaps. Since phonons are quanta of lattice waves,\cite{Ziman_1960_Electrons} it is intuitive that thermal transport will also change in the presence of ordering. Duda \textit{et al.}\cite{Duda_JHT_2011,Duda_JPCM_2011} studied one particular type of order in Si$_\mathrm{0.5}$Ge$_{0.5}$ using nonequilibrium molecular dynamics (NEMD) with a simple Lennard-Jones potential. Baker and Norris\cite{Baker_PRB_2015} later studied both long-range and short-range order in Si$_\mathrm{0.5}$Ge$_{0.5}$ with Stillinger-Weber potentials. We were interested in III-V ternary alloys with compositions away from the 50\%-50\% case. We also  wanted to use EIPs that would give us quantitatively accurate results for TC. It is noteworthy that order in III-V ternary alloy samples is often localized, \textit{i.e.}, it is common to have ``poly-ordering,'' where the sample has one type of order in one region and a different type of order in another.\cite{Kulik_MRSSP_2000,Suzuki_MRS_1999,Shin_MT_2006}

\begin{figure}
	\centering
	\includegraphics[width=0.7\columnwidth]{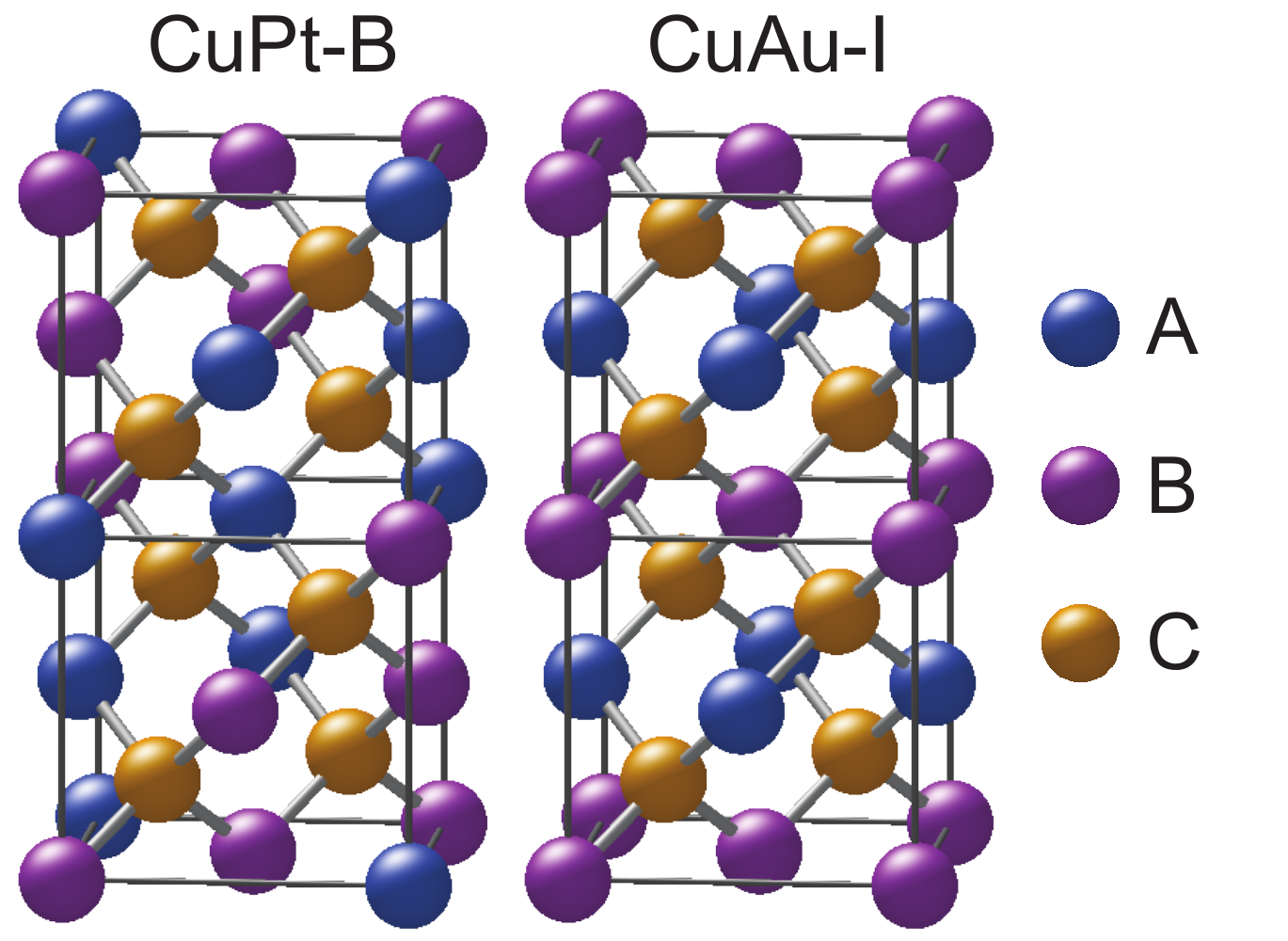}
	\caption{The crystal structure of ordered ternary alloy \ternaryhalf{A}{B}{C} with CuPt-B and CuAu-I ordering.}
	\label{fig:cuptcuau}
\end{figure}

\begin{figure}
	\centering
	\includegraphics[width=0.7\columnwidth]{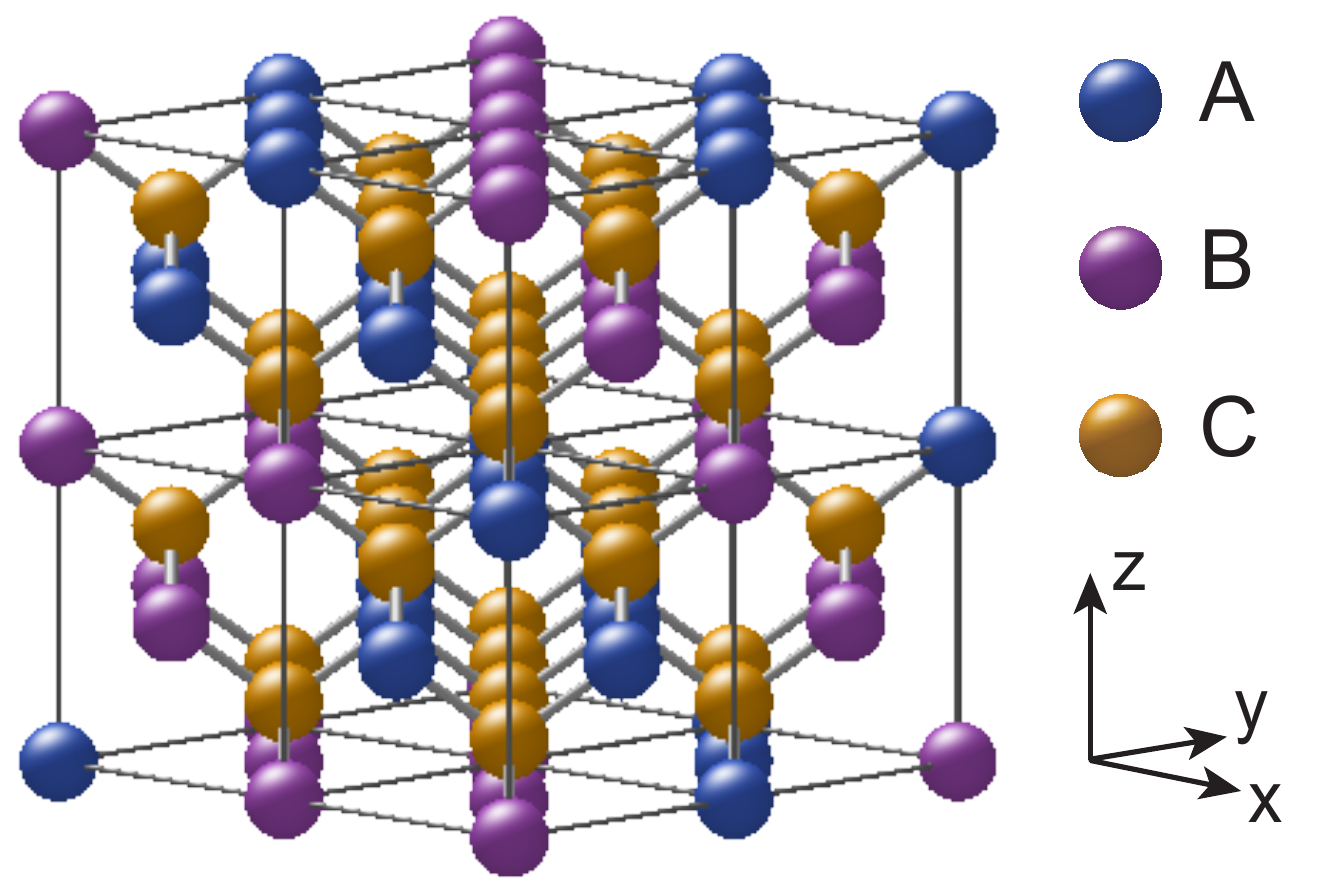}
	\caption{A sample crystal structure of ordered ternary alloy \ternaryhalf{A}{B}{C} with TPA ordering. The 3 planes in a period (\catplane{A}{B}{u}/\catplane{A}{B}{v}/\catplane{A}{B}{w}) have configurations of $u=1,~v=0,$ and $w=0.5$. }
	\label{fig:tpa}
\end{figure}

For the sake of simplicity, in this work we only simulated samples with a single type of order and we applied periodic boundary conditions. Moreover, we focused on three different types of order that are most commonly observed in III-V ternary alloys: the CuPt-B type, the CuAu-I type, and the triple-period-A (TPA) type.\cite{Suzuki_2002_Spontaneous} Both CuPt-B and CuAu-I order yield alloys with the composition \ternaryhalf{A}{B}{C}~within the zinc-blende lattice.  Figure~\ref{fig:cuptcuau} shows the crystal structures for CuPt-B and CuAu-I order. In CuPt-B ordering, A atoms and B atoms take alternate cation planes along the [\={1}11] direction. In CuAu-I ordering, A atoms and B atoms take alternate cation planes along the [100] direction. Note that perfectly ordered alloys do not exist in experiments, so the alternating planes are actually A-rich and B-rich planes. TPA order in its ideal form refers to the case where the cation planes along [111] direction have a repeated pattern involving 3 planes. Therefore, ternary alloy \ternaryx{A}{B}{C} with any composition $x$ can be represented in some TPA ordering where each period has the arrangement of \catplane{A}{B}{u}/\catplane{A}{B}{v}/\catplane{A}{B}{w} and $(u+v+w)/3=x$ (note the three planes in a period cannot be all equal or the triple period collapses). Figure~\ref{fig:tpa} depicts a typical crystal structure of ternary alloy \ternaryhalf{A}{B}{C} where $u=1,~v=0,$ and $w=0.5$.

\subsection{Thermal Conductivity of Alloys Near \ternaryhalf{A}{B}{C} Composition with CuPt-B and CuAu-I order}

We calculated the RT TC of perfectly ordered \ternaryhalf{In}{Ga}{As} and \ternaryhalf{In}{Al}{As} with both CuPt-B and CuAu-I types of order; the results are in Table~\ref{tab:POTC}. The calculated TC for \ternaryhalf{In}{Ga}{As} and \ternaryhalf{In}{Al}{As} with random alloy structures  (with explicit mass) are also shown for comparison. Since there are no experiments on \ternaryhalf{In}{Ga}{As} and \ternaryhalf{In}{Al}{As}, we compared the results with experiments on \InGaAs\cite{Abrahams_JPCS_1959}~and \InAlAs.\cite{Koh_JAP_2009} From Table~\ref{tab:POTC}, we concluded that order of either type significantly increases the TC of both \ternaryhalf{In}{Ga}{As} and \ternaryhalf{In}{Al}{As} alloys compared to the random-alloy case. The CuAu-I type order leads to slightly higher TC than the CuPt-B type order in both alloys. All the simulated TC values from the perfectly ordered structures were higher than experimental measurements, which is intuitive because \InGaAs~and \InAlAs~in experiments are 1) not perfectly ordered and 2) have In content different from 50\%.
\begin{table}
	\begin{ruledtabular}\begin{tabular}{c@{\hspace{1em}}c@{\hspace{1em}}c@{\hspace{1em}}c@{\hspace{1em}}c}
			Material & CuPt-B & CuAu-I & Rand. Alloy & Expt.\\
			\hline
			\ternaryhalf{In}{Ga}{As} & 12.71 & 14.39 & 1.52 & 4.84 (\InGaAs) \\
			\hline
			\ternaryhalf{In}{Al}{As} & 6.488 & 6.472 & 1.25 & 2.68 (\InAlAs)\\
	\end{tabular}\end{ruledtabular}
	\caption{Calculated RT TC for \ternaryhalf{In}{Ga}{As} and \ternaryhalf{In}{Al}{As} with perfectly ordered CuPt-B and CuAu-I structures and for a random alloy. Experimentally measured TC for \InGaAs\cite{Abrahams_JPCS_1959}~and \InAlAs\cite{Koh_JAP_2009} are given for comparison.}
	\label{tab:POTC}
\end{table}

To directly compare the simulated TC with the experimentally measured TC for \InGaAs~and \InAlAs~lattice matched to InP, we randomly replace $3\% ~(2\%)$ of Ga (Al) atoms in the perfectly ordered structure with In atoms to obtain lattice-matched simulation cells. Table~\ref{tab:randLM} shows the TC of lattice-matched \InGaAs~and \InAlAs~from both our simulations and experiments. We see that the TC obtained directly from the lattice-matched simulation cells are still quite large compared to experiment, which could be attributed to the existence of additional disorder that is inevitable in real experimental samples. Here we investigate one type of disorder: randomness. We take the lattice-matched simulation cells and randomly swap an In atom with a Ga or Al atom to create disorder. The amount of disorder is categorized by the percentage of swapped In atoms inside the cell. In Table~\ref{tab:randLM}, the cells with the simulation results are labeled with LM+$d\%$, where LM stands for lattice matched and $d\%$ is the level of disorder. Consistent with the findings in Sec.~\ref{sec:mass}, even little disorder ($<10\%$) severely reduces the alloy TC. Comparing the calculation with experiments, the level of disorder in \InGaAs~ is around 2\% while in \InAlAs~ it is approximately 8\%. Also, as expected, the calculated lattice-matched TC for InGaAs and InAlAs is lower than the corresponding TC calculated from the perfectly ordered structures.
\begin{table}
	\begin{ruledtabular}\begin{tabular}{c@{\hspace{1em}}c@{\hspace{1em}}c@{\hspace{1em}}c@{\hspace{1em}}c@{\hspace{1em}}c@{\hspace{1em}}c@{\hspace{1em}}c}
			& \multicolumn{3}{c}{CuPt-B} & \multicolumn{3}{c}{CuAu-I} &{}\\
			\cmidrule[1pt]{2-4}\cmidrule[1pt]{5-7}
			Material & LM & LM+4\% & LM+8\% & LM & LM+4\% & LM+8\% & Expt.\\
			\hline
			\InGaAs & 5.022 & 3.165 & 2.585 & 5.728 & 3.567 & 2.845 & 4.84 \\
			\hline
			\InAlAs & 3.965 & 3.216 & 2.654 & 4.426 & 3.248 & 2.805 & 2.68 \\
	\end{tabular}\end{ruledtabular}
	\caption{Calculated RT TC for lattice-matched (LM) \InGaAs~and \InAlAs~with various percentages of additional random disorder. Experimentally measured TC values are also listed for comparison.\cite{Abrahams_JPCS_1959,Koh_JAP_2009}}
	\label{tab:randLM}
\end{table}

\subsection{Thermal Conductivity of Alloys with Different Compositions and TPA Order}

\begin{figure}
	\centering
	\includegraphics[width=0.7\columnwidth]{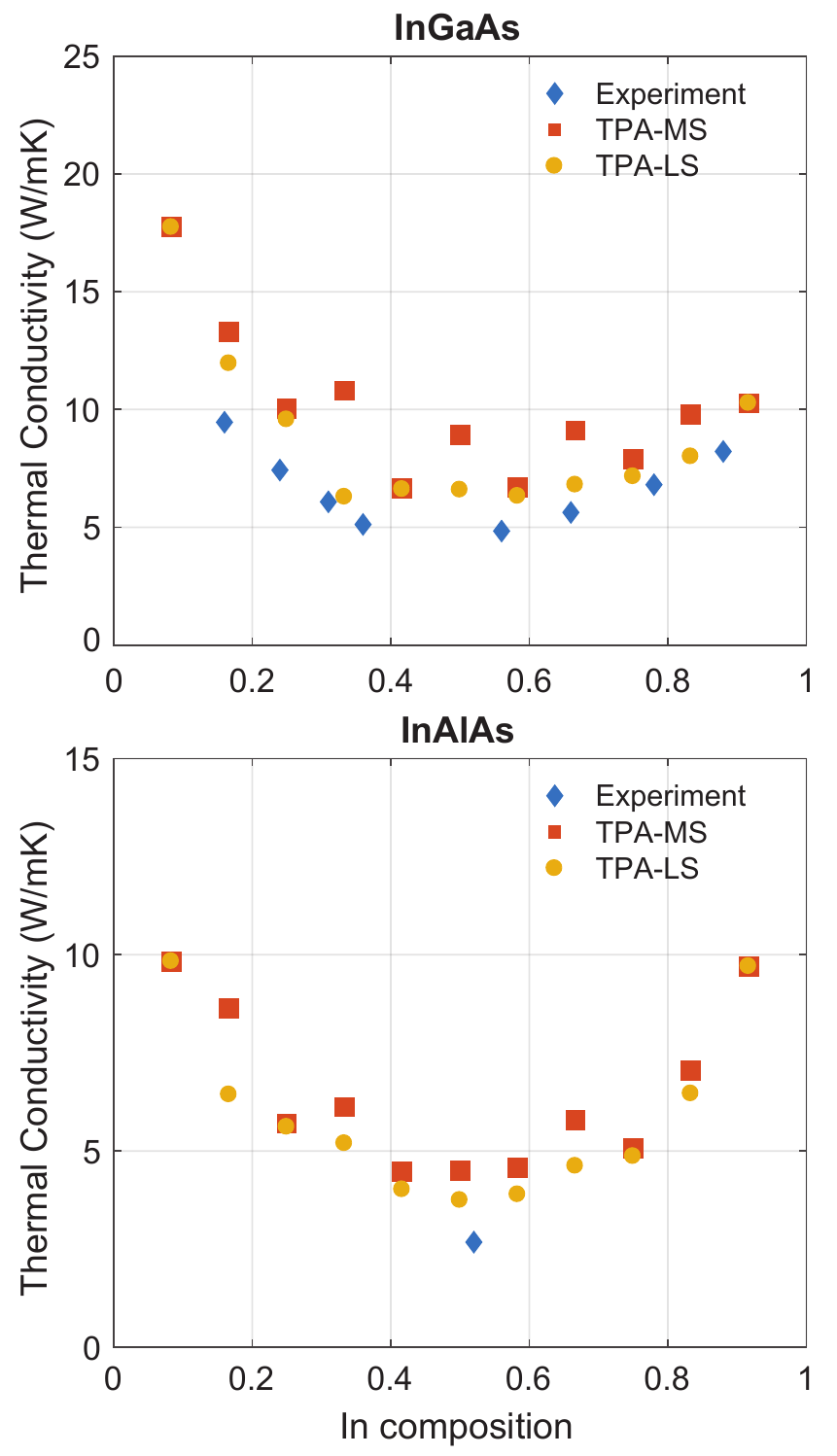}
	\caption{Calculated RT TC of InGaAs and InAlAs with perfect TPA ordering and various In compositions. Red squares and yellow dots show the results from the most symmetric (MS) and least symmetric (LS) structures. Experimental measurements\cite{Abrahams_JPCS_1959,Koh_JAP_2009} are shown in blue diamonds.}
	\label{fig:tpatc}
\end{figure}

To study the TC of ordered \ternaryx{In}{Ga}{As} and \ternaryx{In}{Al}{As} with various $x$ (away from 0.5), we needed to implement TPA order. While it was impossible to consider every feasible arrangement of the TPA order, we simulated the two extreme cases that would likely yield the upper and lower limits of the TC with TPA order for each $x$. The idea is that with more symmetry comes less phonon scattering, and the resulting TC should be higher. Note here we only focused on the perfectly ordered structures, therefore $x$ values were limited to multiples of $\frac{1}{12}$. For each $x$, we generated  two TPA cells: most symmetric (MS) and least symmetric (LS). Figure~\ref{fig:tpatc} shows the TC calculated from the MS cells (red squares), LS cells (yellow dots), and measured in experiment~\cite{Abrahams_JPCS_1959,Koh_JAP_2009} (blue diamonds). The results are consistent with our expectations. The MS TC is higher than the LS TC while both are higher than the experimental data (simulated structures are perfectly ordered, while experimental samples contain disorder). TC values of MS and LS fall on top of each other when $x=\frac{1}{12}$ and $x=\frac{11}{12}$ because the MS and LS structures are equivalent in these cases. The calculated TC follows the general U-shape (TC is lowest when $x$ is close to 0.5 and increases as $x$ approaches 0 or 1), except for a sudden TC jump at certain $x$ values ($x=\frac{n}{12},~n=2,4,6,8,~\text{and}~10$). The jumps exist because these fractions $\frac{n}{12}$ are reducible, which leads to additional symmetry in the system.
\section{Conclusion} \label{sec:conclusion}
In this paper, we studied the thermal conductivity of III-V ternary alloys, in particular \ternaryx{In}{Ga}{As}~and \ternaryx{In}{Al}{As}. Using equilibrium molecular dynamics, we investigated how the mass difference between cation atoms and the arrangement of cations affect thermal transport. Optimized Able-Tersoff EIPs were employed, owing to their demonstrated success in describing III-V binary materials and their nearest-neighbor cutoffs that lend themselves to application in alloys.

We first optimized the Able-Tersoff potentials by matching the calculated phonon dispersion to experiments. We also matched the calculated temperature-dependent TC to experiment so the EIPs would capture higher orders in phonon--phonon interaction that are key in thermal transport. The quantum correction to the temperature was also accounted for in the simulations.

The optimized potentials were used to describe ternary alloys with both random and ordered structures. For random alloys, we compared the cases where atom masses were explicitly considered versus where they were all replaced with the averaged VCA mass, as is commonly done within the Boltzmann transport framework. The results showed that explicit atomic mass drastically reduces the TC of ternary alloys. The larger the mass difference between the cations in the alloy, the larger the discrepancy between explicit-mass and averaged-mass VCA  TC results. We concluded that, when cation masses differ a great deal (as is the case for \ternaryx{In}{Al}{As}), it is essential to include atomic masses explicitly and any calculation that relies on VCA is likely inaccurate.

Moreover, measured thermal conductivities of ternary alloys are higher than calculations with either explicit or VCA mass and random positioning of cations, which led us to look at longer-range order in alloys. We considered  perfectly ordered \ternaryhalf{In}{Ga}{As}~and \ternaryhalf{In}{Al}{As}~with both CuPt-B and CuAu-I types of order and the corresponding lattice-matched alloys \InGaAs~and \InAlAs. Order in ternary alloys considerably raises TC. By adding random disorder to the lattice-matched alloys \InGaAs~and \InAlAs, we found that experimental results could be reproduced with levels of disorder close to 2\% in \InGaAs~and 8\% in \InAlAs.

We also studied perfectly ordered TPA alloys \ternaryx{In}{Ga}{As}~and \ternaryx{In}{Al}{As}~with various compositions $x$. We found that more symmetry in the alloy led to higher TC, while the alloys with the least symmetry still yielded higher TC than experiments, indicating the existence of disorder in experiment.

In conclusion, in modeling thermal transport in III-V ternary alloys, it is crucial to include both the explicit masses of atoms and the effects of  long-range order. The measured TC for III-V ternary alloys is likely a result of the competition between the two: reduction in TC stemming from mass-difference scattering associated with random disorder and an increase in TC associated with order in the alloy structure. These notions should be incorporated into other techniques for calculating thermal transport in alloys, and highlight the importance of critically evaluating the range of validity of even very common approximations, such as the VCA.

\begin{acknowledgments}
The authors thank Luke Mawst for bringing the issue of long-range order in III-V alloys to our attention. We are also grateful to Dan Botez, Mark Eriksson, Jeremy Kirch, Gabriel Jaffe, and Colin Boyle for useful discussions. This work was supported by the the U.S. Department of Energy, Office of Science (Basic Energy Sciences, Division of Materials Sciences and Engineering, Physical Behavior of Materials Program) under Award No. DE-SC0008712. The simulation work was performed using the compute resources and assistance of the UW-Madison Center For High Throughput Computing (CHTC) in the Department of Computer Sciences.
\end{acknowledgments}


%

\end{document}